\documentstyle[aps,prl,twocolumn]{revtex}

\begin{document}
\draft
\title{Controlling Quantum Rotation With Light}
\author{I. Sh. Averbukh*, R. Arvieu$\dagger $, and M. Leibscher*}
\address{$\ast $ Department of Chemical Physics, The Weizmann Institute of Science,
Rehovot 76100, Israel\\
$\dagger $ Institut des Sciences Nucl\'{e}aires, F 38026 Grenoble Cedex,
France\\
e-mail: Ilya.Averbukh@weizmann.ac.il}
\maketitle

\begin{abstract}
Semiclassical catastrophes in the dynamics of a quantum rotor (molecule)
driven by a strong time-varying field are considered. We show that for
strong enough fields, a sharp peak in the rotor angular distribution can be
achieved via time-domain focusing phenomenon, followed by the formation of
angular rainbows and glory-like angular structures. Several scenarios
leading to the enhanced angular squeezing are proposed that use specially
designed and optimized sequences of pulses. The predicted effects can be
observed in many processes, ranging from molecular alignment (orientation)
by laser fields to heavy-ion collisions, and the squeezing of cold atoms in
a pulsed optical lattice.
\end{abstract}

\pacs{PACS numbers: 42.50.-p, 42.50.Vk, 32.80.Pj, 33.80.-b}

\narrowtext

\section{Introduction}

Driven rotor is a standard model in classical and quantum nonlinear dynamics
studies\cite{Haake}. Increasing interest in the problem has arisen because
of recent atom optics realization of the quantum $\delta $-kicked rotor \cite
{Raizen,Ammann}, and novel experiments on molecule orientation (alignment)
by strong laser fields\cite{Normand,Dietrich,Mathur,Corkum}. A strong enough
laser field creates the so-called pendular molecular states \cite
{Zon,Friedrich1,Felker96} that are hybrids of field-free rotor eigenstates.
By adiabatically turning on the laser field, it is possible to trap a
molecule in the ground pendular state, thus leading to molecular alignment.
The only way to reach a considerable degree of alignment in this approach is
by increasing the intensity of the field. However, many applications may
require only a transient molecular alignment (orientation), where the
molecular angular distribution becomes extremely squeezed at some
predetermined moment of time. It is well known that a physically related
problem of squeezed states generation in a harmonic system may be solved by
a proper time-modulation of the driving force (parametric resonance
excitation). Behavior of a rotor in general, strong, time-varying fields is
a much less-studied problem, although it is understood that the
long-persisting beats in the molecular angular distribution may be induced
by short laser pulses \cite{Lin1,Fonda,Felker,Tamar1,Tamar2,Ortigoso,Cai}.

In the present paper, we analyze generic features in the dynamics of a
quantum rotor driven by strong pulses, and present a strategy for efficient
squeezing of the rotor angular distribution by a sequence of pulses of
moderate intensity. The results of our research \ are related to a number of
physical processes, ranging from molecular alignment (orientation) by laser
fields to heavy-ion collisions, and the trapping of cold atoms by a standing
light wave. The paper is organized as following. Section II discusses
semiclassical catastrophes in the dynamics of a driven quantum rotor with
the help of a simple two-dimensional model \cite{rotor}. Section III applies
these generic results to a thermal system of cold atoms driven by a pulsed
optical lattice \cite{Monika}. Section IV focuses on 3D effects important in
the processes of molecular alignment and orientation by strong laser pulses 
\cite{3D}.

\section{Two-dimensional rotor}

For the sake of clarity, we start with the simplest model of a
two-dimensional rotor described by the time-dependent Hamiltonian

\begin{equation}
\widehat{H}=\frac{\widehat{L}^{2}}{2I}+V(\theta ,t)  \label{hamiltonian2}
\end{equation}
where $\widehat{L}$ is the angular momentum operator, and $I$ is the
momentum of inertia of the rotor. \ This model contains most of the physics
we want to present here. Some important additional effects that appear in 3D
will be discussed below in Section IV. For a ''linear 2D molecule'' having a
permanent dipole moment $\mu ,$ and driven by a linearly polarized field,
the interaction potential is 
\begin{equation}
V(\theta ,t)=-\mu {\cal E}(t)\cos (\theta )  \label{permanent}
\end{equation}
where ${\cal E}(t)$ is the field amplitude (i. e., of a half-cycle pulse),
and $\theta $ is the polar angle between the molecular axis and the field
direction. In the absence of interaction with a permanent dipole moment, the
external field may couple with induced anisotropic molecular polarization.
This interaction (being averaged over fast optical oscillations) may result
in the interaction potential proportional to $\cos ^{2}(\theta )$ (see Eq.(%
\ref{polarization}) in Section IV) \cite{Boyd}. Although these two forms of $%
V(\theta ,t)$ may lead to different physical consequences (i. e.,
orientation vs alignment), the effects we will present are more or less
insensitive to the choice of interaction. Therefore, we prefer to start with
the more simple situation described by Eq.(\ref{permanent}).

By introducing dimensionless time $\tau =t\hbar /I,$ and interaction
strength $\varepsilon =\mu {\cal E}(t)I/\hbar ^{2},$ the Hamiltonian can be
written as 
\[
\widehat{H}=-\frac{1}{2}\frac{\partial ^{2}}{\partial \theta ^{2}}%
-\varepsilon (\tau )\cos (\theta ) 
\]
The wave function of the system can be expanded in the eigenfunctions of a
free rotor 
\[
\Psi (\theta ,\tau )=\frac{1}{\sqrt{2\pi }}\sum_{n=-\infty }^{+\infty
}c_{n}(\tau )e^{in\theta } 
\]
In the absence of the field, the wave function takes the form 
\begin{equation}
\Psi (\theta ,\tau )=\frac{1}{\sqrt{2\pi }}\sum_{n=-\infty }^{+\infty
}c_{n}(0)e^{-in^{2}\tau /2+in\theta }  \label{wavefun}
\end{equation}
Despite a simple form of Eq.(\ref{wavefun}), the wave function exhibits
extremely rich space-time dynamics. In particular, it shows periodic
behavior in time with the period $T_{rev}=4\pi $ (full revival) and a number
of fractional revivals at $\tau =p/sT_{rev}$ ($p$ and $s$ are mutually prime
numbers) \cite{fr}. An analytical solution valid for a general
time-dependent field is unknown even for this simplest model. Much effort
has been devoted to the case of extremely short field pulses ($\delta -$%
kicks) (see, e. g. \cite{Haake}, and references therein). In general, as a
result of a single kick applied to the rotor at $\tau =\tau _{k},$ the
coefficients $c_{n}$ transform as 
\begin{equation}
c_{n}(\tau _{k}+0)=\sum\limits_{m=-\infty }^{+\infty
}i^{n-m}J_{n-m}(P)c_{m}(\tau _{k}-0),  \label{trans}
\end{equation}
where 
\[
P=\int\limits_{-\infty }^{+\infty }\varepsilon (\tau )d\tau , 
\]
and $J_{n}(P)$ is the Bessel function of the $n$th order. The result of
multiple kicks applied at different times can be obtained by combining
transformations (\ref{trans}) after each kick with a free evolution
according to Eq.(\ref{wavefun}) between the kicks. If the kicks are applied
periodically to the system with the period $T_{rev},$ the system does not
show chaotic behavior, and the energy accumulates quadratically with time
(the so-called ''quantum resonance'' \cite{Como,Izrailev1}). It is,
therefore, quite natural to examine potential accumulation of angular
squeezing of the rotor wave function under the ''quantum resonance''
excitation. In this case, because of the exact quantum revivals at the
free-evolution stages, the effect of $N$ kicks of a magnitude $P$ is
equivalent to the action of a single strong pulse of strength $NP$ (see, e.
g. \cite{Como}). In Figure 1, we show numerically calculated time evolution
of the probability density $|\Psi (\theta ,\tau )|^{2}$ after a relatively
strong kick of a magnitude $P=85$ applied at $\tau =0.$ Initially the rotor
was in the ground $s$-state ($c_{n}(0)=\delta _{n0})$. For the chosen values
of $\tau $, several distinct phenomena can be seen in these plots. First of
all, the wave function shows an extreme narrowing in the region of small $%
\theta $ after some delay following the kick [Fig. 1 (b)]. The physics of
this effect may be understood with the help of the following semiclassical
arguments. Consider an ensemble of randomly oriented classical rotors
subject to a kick. The angular velocity of a rotor located at the angle $%
\theta $ is 
\begin{equation}
\omega (\theta )=-P\sin (\theta )  \label{velocity}
\end{equation}
just after the kick, assuming negligible initial velocity. For rotors from
the region of small $\theta <<1,$ the acquired velocity is linearly
proportional to their initial angle, so that all of them arrive at the focal
point $\theta =0$ at the same time 
\begin{equation}
\tau _{f}=1/P.  \label{focal time}
\end{equation}
This phenomenon is quite similar to the focusing of light rays by a thin
optical lens. For $P>>1$, the shape of the distribution at the focusing time 
$\tau _{f}$ is dictated by the aberration mechanism (deviation of the $\cos
(\theta )$ potential from the parabolic one), and it is $P$-independent. We
consider the orientation factor $O=<1-\cos (\theta )>$ (where angular
brackets mean averaging over the state of rotor) as a measure of the rotor
orientation. For large enough $P,$ the time-dependent orientation factor
(for the initial $s$-state) may be easily estimated by averaging over the
initially uniform classical ensemble of rotors having the velocity
distribution of Eq. (\ref{velocity}): $O(\tau )=1-J_{1}(P\tau )$. Here $%
J_{1}(x)$ is Bessel function of the first order. The minimal value $O\approx
0.418$ of the orientation factor is, in fact, achieved in the post-focusing
regime, at $\tau \approx 1.84\tau _{f}$. As seen in Fig. 1 (c), a new
phenomenon can be observed in the angular probability distribution just
after the focusing. Sharp singular-like features are formed in the
distribution, which are moving with time. Each of these features has a
typical asymmetric shape, with pronounced oscillations on one side and an
abrupt drop down on the other side. Again, the origin of this effect can be
traced in the time evolution of a classical ensemble of initially motionless
rotors.

After a kick applied at $\tau =0$, the motion of the rotors is described by 
\begin{equation}
\theta =\theta _{0}-P\sin (\theta _{0})\tau \text{ \ (mod }2\pi \text{)}
\label{map}
\end{equation}
where $\theta _{0}$ is the initial angle. For $\tau <\tau _{f}$ \ Eq.(\ref
{map}) represents a one-to-one mapping $\theta (\theta _{0})$ [see Fig. 2
(a)]. At $\tau =\tau _{f}$ the curve $\theta (\theta _{0})$ touches the
horizontal axis [Fig. 2 (b)]. At $\tau >\tau _{f}$ the angle $\theta _{0}$
becomes a multi-valued function of $\theta $ [Figs. 2 (c), (d)]. The
classical time-dependent angular distribution function of the ensemble is
given by 
\begin{equation}
f(\theta ,\tau )=\sum\limits_{a}\frac{f(\theta _{0}^{a},\tau =0)}{|d\theta
/d\theta _{0}^{a}|}  \label{dist}
\end{equation}
The summation in Eq. (\ref{dist}) is performed over all possible branches of
the function $\theta _{0}(\theta )$ defined by Eq.(\ref{map}). It follows
immediately from Eq.(\ref{dist}) that even for a smooth initial
distribution, $f(\theta ,\tau )$ may exhibit a singular behavior near the
angles where $d\theta /d\theta _{0}^{a}\rightarrow 0.$ The quantum nature of
the rotor motion replaces the classical singularities by sharp maxima in the
probability distribution with the Airy-like shape typical to rainbow
phenomena. Indeed, this effect is similar to the formation of caustics in
the wave optics \cite{Kravtsov}, and rainbow-type scattering in optics and
quantum mechanics \cite{Ford,Berry,Bosanac}. We should stress, however, that
the long-time asymptotic regimes are radically different for the classical
and truly quantum motion of the rotor. Thus, contrary to the classical
limit, in which the caustics exist forever, they gradually disappear in the
quantum case because of the overall decay of the initial rotational wave
packet. On even longer time scale, another quantum phenomenon can be seen,
namely revivals and fractional revivals of the initial classical-like
motion. Figs. 1 (e)-(i) show several examples of fractional foci and
rainbows in the angular distribution, which is a purely quantum effect.

As we have demonstrated, a mere application of $\delta -$kicks at the
condition of ''quantum resonance'' does not lead to accumulated angular
squeezing, and the orientation is saturated at some finite asymptotic level.
Here we suggest an excitation scheme that exhibits the desired accumulation
property. As previously mentioned, the wave function of the rotor reaches
the state of the maximal orientation (i. e., minimal $O$ value) after a
certain delay $\Delta \tau _{1}$ following the application of the first kick
at $\tau =\tau _{1}=0$. We suggest to apply the second kick at $\tau
_{2}=\Delta \tau _{1}.$ Immediately after the second kick, the system will
keep the same probability density distribution. On the other hand, $\tau
=\tau _{2}$ will no longer be a stationary point for $O(\tau )=<1-\cos
(\theta )>(\tau ).$ The orientation factor $O(\tau )$, and its derivative
are continuous and periodic functions of time in the course of a free
evolution. Therefore, $O(\tau )$ will reach a new minimum at some point $%
\tau _{2}+\Delta \tau _{2}$ in the interval $[\tau _{2},\tau _{2}+T_{rev}]$.
Clearly, the new minimal value of the orientation factor is smaller than the
previous one. By continuing this way, we will apply short kicks at iterative
time instants $\tau _{k+1}=\tau _{k}+\Delta \tau _{k}.$ By construction of
this pulse sequence, the squeezing effect will accumulate with time, in
contrast to the ''quantum resonance'' excitation. This is demonstrated by
Fig. 3, which shows calculated sequences $\{\Delta \tau _{k}\}$ and $%
\{<1-\cos(\theta )>(\tau _{k})\}$ for a rotor initially in the $s$-state and
being kicked by pulses with $P=3$. The logarithm of the orientation factor
gradually decreases, without any sign of saturation.

At the stage of a well-developed squeezing ($O<<1$), the $\cos (\theta )$
-potential may be approximated by a parabolic one. It can be easily shown
(both classically and quantum mechanically) that in this limit our strategy
provides the following recurrent relationships for the time intervals $%
\Delta \tau _{k},$ successive values of the angular variance $u_{k}=<\theta
^{2}>_{k}\approx 2O_{k}$ and normalized variance of the angular velocity $%
w_{k}=<(-id/d\theta )^{2}>_{k}/P^{2}$ : 
\begin{eqnarray}
\Delta \tau _{k} &=&\frac{u_{k}}{u_{k}+w_{k}}  \nonumber \\
u_{k+1} &=&u_{k}-\frac{u_{k}^{2}}{u_{k}+w_{k}}  \label{system} \\
w_{k+1} &=&w_{k}+u_{k}  \nonumber
\end{eqnarray}
For large $k$, the last two finite-difference equations may be replaced by a
system of coupled differential equations. The latter has an exact solution
providing the following asymptotics: $<\theta ^{2}>_{k}\propto 1/\sqrt{k}$
and $\Delta \tau _{k}\propto 1/k$. This result is in good agreement with the
numerically observed power-laws behavior of the graphs 3 (a) and 3 (b), and
it describes correctly their slopes at $k>>1$. We note, that in contrast to
the wave optics (in which the size of the focal spot is diffraction
limited), our system may be, in principle, ''unlimitedly'' squeezed in
angle. We also note that a quasi-periodic sequence of kicks applied at $\tau
_{k+1}=\tau _{k}+\Delta \tau _{k}+T_{rev}$ provides the same squeezing
scenario for a quantum rotor. The introduction of the $T_{rev}$-shift
between pulses may be useful in the practical realizations of the scenario
to avoid the overlap between short excitation pulses of a finite duration.

\section{Squeezing of atoms in a pulsed optical lattice}

In the present Section we apply the above results to another well-known
system : cold atoms interacting with a pulsed optical lattice \cite{Monika}.
Optical lattices are periodic potentials for neutral atoms induced by
standing light waves formed by counter-propagating laser beams. When these
waves are detuned from any atomic resonance, the ac Stark shift of the
ground atomic state leads to a conservative periodic potential with spatial
period $\lambda /2,$ half the laser wavelength (for a review, see, e. g. 
\cite{Jessen}). Such lattices present a convenient model systems for solid
state physics and nonlinear dynamics studies. In contrast to traditional
solid state objects, the parameters of optical lattices (i. e. lattice
constant, potential well depth, etc.) are easily controllable. Many fine
phenomena that were long discussed in solid state physics, have been
recently observed in corresponding atom optics systems. Bloch oscillations 
\cite{Dahan}, or the Wannier-Stark ladder \cite{Wilkinson,Niu} are only few
examples to mention. Time-modulation of the frequency and intensity of the
constituent laser beams provide tools for effective modeling of numerous
time-dependent nonlinear phenomena. Since the initial proposal \cite{Graham}%
, and first pioneering experiments on atom optics realization of the $\delta 
$-kicked quantum rotor \cite{Raizen}, cold atoms in optical lattices provide
also new grounds for experiments on quantum chaos.

We describe atoms as two-level systems with transition frequency $\omega
_{0} $, interacting with a standing optical wave that is linearly polarized
and has the frequency of $\omega _{l}$. If the detuning $\Delta _{l}=\omega
_{0}-\omega _{l}$ is large compared to the relaxation rate of the excited
atomic state, the internal structure of the atoms can be neglected and they
can be regarded as point-like particles. In this approximation, the
Hamiltonian for the atomic motion is 
\begin{equation}
H(x,p,t)=\frac{p^{2}}{2m}-V(t)\cos \left( 2k_{l}x\right) ,  \label{ham1}
\end{equation}
where $m$ is the atomic mass, $k_{l}=\omega _{l}/c$ is the wave number of
the standing wave. The depth of the potential produced by the standing wave
is $V(t)=\hbar \Omega (t)^{2}/8\Delta _{l}$, where $\Omega
(t)=2|d_{z}E(t)|/\hbar $ is the Rabi frequency, $\vec{d}$ is the atomic
dipole moment and $E(t)$ is the time-dependent amplitude of the light field.
In the case of rather short laser pulses, this Hamiltonian corresponds to
that of the two-dimensional $\delta $-kicked rotor considered in the
previous Section, with $\theta =$ $2k_{l}x$. In accordance with the previous
discussion, many aspects of the dynamics of this system can be explained
with semiclassical arguments. Here we provide results for a classical
description of atoms in a pulsed optical lattice, with thermal effects taken
into account.

We performed a Monte-Carlo simulation of the dynamics of an ensemble of $%
\delta $-kicked particles which were initially uniformly distributed in
space and had a thermal momentum distribution. Figure 4 shows the spatial
distribution of atoms at different times after a single kick. In the upper
row, the initial temperature of the ensemble is zero, and we can observe
focusing [Fig. 4(a)] and formation of caustics [Fig. 4(b)] as it was
described in Section II. In Fig. 4(c) and (d), the average thermal energy of
the initial ensemble is comparable with the energy supplied by a kick. As a
result, instead of sharp peaks in the spatial distribution (like in Figure 4
(a) and (b)), we observe some broader spatial structures that are still
reminiscent of the focusing phenomenon and the rainbow effect.

We examined the accumulative squeezing approach in application to the above
system. The angular localization factor $O=<1-\cos \left( 2k_{l}x\right) >$
introduced in Section II describes now the spatial width of atomic groups
localized in the minima of the light-induced potential. In the classical
limit, the localization factor after applying $n$ kicks is given by 
\begin{equation}
O(t_{n})=1-\frac{1}{2\pi }\int_{-\infty }^{\infty }d\omega _{0}\int_{-\pi
}^{\pi }d\theta _{0}\rho (\omega _{0})\cos \theta _{n},  \label{loctn}
\end{equation}
where $\theta _{n}$ is determined by 
\begin{equation}
\theta _{n}=\theta _{n-1}+\Delta \tau _{n}\left( \omega
_{0}-\sum_{i=0}^{n-1}\sin \theta _{i}\right) .
\end{equation}
Here, $\omega _{0}=2k_{l}v_{0}$, where $v_{0}$ is the initial velocity of an
atom, and $\rho (\omega _{0})$ describes the thermal distribution of the
velocities. Figure 5 displays the amount of spatial squeezing for the series
of 100 kicks for different initial temperatures. It can be seen, that after
the first few kicks the localization factor demonstrates a negative power
dependence as a function of the kick number (a straight line in the double
logarithmic scale). Although, the system demonstrates a reduced squeezing
for higher initial temperatures, the slope of all of the curves in Figure 5
is the same after the first several kicks, in full agreement with the
general arguments of the previous Section (see Eqs.(\ref{system}).
Therefore, the accumulative squeezing scenario may be an effective and
regular strategy for atomic localization even at finite temperatures.

However, this does not mean that the accumulative squeezing is the only one
(or the most effective) squeezing strategy. We have studied the best
localization that can be achieved with {\em a given number} of identical $%
\delta $-kicks, by minimizing the localization factor Eq.(\ref{loctn}) with
respect to the delay times $\Delta \tau _{n}$ between the kicks. For
clarity, we present here only the results for zero initial temperature of \
the atoms. Table 1 shows the best values of the localization factor found
for up to five kicks, and compares them with the results of the accumulative
squeezing strategy with the same number of kicks. While the maximal atomic
localization that can be achieved with two pulses is almost the same for
accumulative squeezing and for the optimal sequence of two pulses, the
optimized results for three and more pulses are much better.

For illustration, we choose the sequence of four optimal pulses to visualize
the dynamics behind the localization process. Figure 6 (a) shows the spatial
distribution of the ensemble of atoms at the time of arrival of the second
pulse. Note that the second pulse is not applied at the time of the maximal
localization. On the contrary, the optimized procedure finds it favorable to
wait after the focusing event, until the distribution becomes rather broad.
The optimal four-pulse sequence we found requires applying the third and the
forth pulses simultaneously, thus producing an effective ''double pulse''.
The spatial distribution at the time of the combined third and fourth pulses
can be seen in Figure 6 (b). The distribution is only slightly more
localized than in Fig. 6(a), and only the last pulse (with the double
strength) squeezes the ensemble at the time of the maximal localization,
thus bringing most of the atoms to the optical lattice minima [see Fig. 6
(c)].

\section{Orientation and alignment of a 3D rotor}

Under certain conditions, the process of molecular orientation (or
alignment) by laser fields can be described by a strongly driven
three-dimensional rigid rotor model. Although many features in the dynamics
of 3D rotors are similar to those already discussed for the two-dimensional
case, there are two principal differences that we want to emphasize. The
first one may be traced in the evolution of a classical ensemble of
3D-rotors being initially at zero temperature. The probability of finding a
rotor (driven by a linearly polarized field) at a certain solid angle
element $\sin \theta d\theta $ is determined by the initial distribution
function, $f_{0}(\theta _{0})$ as follows: 
\begin{equation}
f(\theta )\sin \theta d\theta =f_{0}(\theta _{0})\sin \theta _{0}d\theta
_{0}.
\end{equation}
Therefore, the probability density of finding a rotor at the angle $\theta $
at some latter time is 
\begin{equation}
f(\theta )=\sum_{a}\frac{f_{0}(\theta _{0}^{a})\sin \theta _{0}^{a}}{\left|
d\theta /d\theta _{0}^{a}\right| \sin \theta },
\end{equation}
where the summation is done over all branches of the $\theta _{0}(\theta )$
function (see Section II). The probability density $f(\theta )$ has a
singularity if one of the factors in the denominator is zero. As in the 2D
case, the zeroes of the fraction $\left| d\theta /d\theta _{0}^{a}\right| $
give rise to the formation of the rainbow-like structures. But in addition,
the geometrical factor $\sin \theta $ can get zero too, which leads to the
additional singularities at $\theta =0$ and $\theta =\pi .$ This kind of
singularities are responsible for the formation of the {\em corona }and {\em %
glory} effects in the optical and quantum-mechanical scattering \cite
{Ford,Berry}.

The second difference appears at finite initial temperature of the ensemble
of 3D rotors. Because of the conservation of the angular momentum projection
onto the field polarization direction, an effective repulsive centrifugal
force appears that prevents the rotors from reaching the exact $\theta =0$
and $\theta =\pi $ orientations. As a result, two {\em holes} in the angular
distribution of a driven 3D thermal ensemble should be always present at $%
\theta =0$ and $\theta =\pi $ \cite{Zon1}$.$

We present below the results of Monte-Carlo simulation of the dynamics of a
classical ensemble of three-dimensional rotors (linear molecules) driven by
a linearly polarized time-dependent field. The corresponding Hamiltonian is 
\begin{equation}
H=\frac{1}{2mr^{2}}\left( p_{\theta }^{2}+\frac{p_{\phi }^{2}}{\sin
^{2}\theta }\right) -\mu {\cal E}(t)r\cos \theta ,
\end{equation}
where $m$ is the reduced mass of the molecule, $r$ is the (fixed) distance
between the atoms, $\mu $ is permanent dipole moment, $\theta $ and $\phi $
are Euler angles, and $p_{\theta }$ and $p_{\phi }$ are related canonical
momenta. For the driving field described by a $\delta -$ pulse, the
equations of motion can be easily integrated (see \cite{goldstein}).

At zero initial temperature, $p_{\theta _{0}}=0$ and $p_{\phi _{0}}=0$, and
the angle $\phi (t)$ is an additional constant of motion. In this case, the
dynamics of the system is reduced to that of a two-dimensional rotor, beside
the effect of the geometrical factor described above. Figure 7 demonstrates
the time evolution of the angular distribution of an ensemble with finite
initial temperature. Here, the probability of finding a rotor inside the
solid angle element $\sin \theta d\theta d\phi $ is plotted on a sphere. Red
color corresponds to the high probability density while the blue color
presents a low probability. Initial isotropic ensemble corresponds to an
uniform solid angle distribution [see Fig. 7 (a)]. In Figure 7 (b), the
distribution is plotted at the ''focal time'' defined according to Eq.(\ref
{focal time}). In the two-dimensional case considered in the previous
Sections, the distribution at the focal time is characterized by a sharp
peak at $\theta =0$ and a broad background for larger values of $\theta $.
In the three-dimensional case, the distribution has an additional sharp dip
at $\theta =0$, as discussed above. Figures 7 (c) and (d) show the angular
distribution at $t>t_{f}$. In Fig. 7 (c) the formation of the ''{\em corona}%
'' around $\theta =0$ can be seen. In addition, we can observe a ring of
relatively high probability moving from the north pole to the south pole of
the sphere. This ring is a three-dimensional analog of the rainbow structure
modified by the thermal effects. After the ring arrives at the south pole, a
singular feature appears around $\theta =\pi $, with a hole in the center
caused the repulsive centrifugal force [Fig. 7 (d)]. The formation of this
sharp and robust structure in the angular distribution is analogous to the
glory effect in the wave optics.

All the described phenomena, namely, focusing, caustics creation, and
accumulating squeezing are not specific for the simplest models considered
above, but are rather common features that can be observed under general
conditions of a strong excitation of the quantum rotor. Indeed, for strong
enough driving field one may neglect the initial rotational energy stored in
the rotor, and use the above quasiclassical ideas to describe its dynamics.
The role of the initial state of the rotor (even thermally averaged) is
reduced to the formation of a frozen classical-like initial angular
distribution of the rotational ensemble. For example, for the dipole-type
interaction, Eq. (\ref{permanent}), the angular focusing may be analyzed by
considering classical dynamics \ at small angles ($\theta \ll 1$): 
\begin{equation}
\left[ \frac{d^{2}}{d\tau ^{2}}+\epsilon (\tau )\right] \theta =0.
\label{multifoc}
\end{equation}
Here, $\tau =t\hbar /I$, and $\epsilon =\mu {\cal E}(t)I/\hbar ^{2}$, and
the initial condition is $\theta (\tau _{0})=\theta _{0},$ $d\theta (\tau
_{0})/d\tau =0$ (where $\tau _{0}$ is any moment in the past before the
beginning of the pulse). Focusing time $\tau _{f}$ is defined by $\theta
(\tau _{f})$ $=0$. Because of the linearity of Eq.(\ref{multifoc}), the
position of the focusing times, and number of focusing events do not depend
on $\theta _{0},$ but are determined only by the properties of the pulse $%
\epsilon (\tau )$. This boundary problem may be solved analytically in
several special cases only (for $\delta $-pulse considered above, or for a
step-like $\epsilon (\tau )$ dependence). In general, determination of
focusing times requires numerical solution of Eq.(\ref{multifoc}). Based on
the previous analysis, we expect that a new angular rainbow appears
immediately after each focusing event.

Figure 8 shows numerically calculated time-evolution of a three-dimensional 
{\em quantum }rotor (linear molecule) coupled to the external laser field
via the anisotropic polarization interaction. The time averaged value of
this interaction is (see, e. g.\cite{Boyd,Friedrich1}) 
\begin{equation}
V(\theta ,t)=-1/4{\cal E}^{2}(t)[(\alpha _{\Vert }-\alpha _{\bot })\cos
^{2}(\theta )+\alpha _{\bot }]  \label{polarization}
\end{equation}
Here $\alpha _{\Vert }$ and $\alpha _{\bot }$ are the components of the
polarizability, parallel and perpendicular to the molecular axis, and ${\cal %
E}(t)$ is the {\em envelope} of the laser pulse. In this case there are two
opposite equilibrium directions $\theta =0,\pi $, so the field aligns the
molecules but not orients them. The applied pulse has a Gaussian shape of a
finite duration (i. e. it is not a $\delta $-kick). As seen from Fig. 8,
both focusing and caustics formation occur in the system, similar to the
previous consideration, with the differences attributed to another symmetry
of the problem. For instance, the angular rainbows are made of moving rings
located symmetrically with respect to the equatorial plane. Note, that the
glory and corona are not seen because of the $\sin (\theta )$ factor
incorporated into the distribution function shown in Fig. 8.

\section{Conclusions}

The predicted effects may be observed in a wide range of systems with
strongly driven rotational degrees of freedom. Possible examples range from
heavy-ion collisions (when highly excited wave packets of nuclear rotational
states are produced \cite{Fonda}) to molecules subject to strong laser
pulses, and cold atoms trapped by standing light waves. The spectacular
features described in this paper may be observed in the spatial distribution
of an atomic ensemble driven by pulsed optical lattices. Recently, the
accumulated squeezing scenario \cite{rotor} has been realized experimentally
in this system \cite{experiment}. Moreover, the related squeezing approaches
may find application in atom lithography of ultra-high resolution \cite
{patent,lith}. In the case of molecules, the considered effects may reveal
themselves in the angular distribution of fragments produced by intense
laser-field molecular interaction. The most direct evidence can be achieved
in a two-pulse experiment, in which the first strong non-resonant pulse
attempts to orient (align) the molecular ensemble, while the second short
delayed pulse creates fragment ions.

{\bf Figure Captions:}

\bigskip

Table 1. The table shows the minimal value of the localization factor that
can be achieved with a fixed number of kicks using accumulative squeezing
scenario ($O_{acc}$) and optimized sequence of pulses ($O_{opt}$).

\bigskip

Fig. 1. Angular distribution of a quantum rotor excited by a strong $\delta $%
-kick ($P=85$). The graphs correspond to (a) $\tau =0.5\tau _{f},$ (b)$\
\tau =\tau _{f},$ (c) $\tau =2\tau _{f}$, (d) $\tau =\tau _{f}+T_{rev}/2,$
(e) $\tau =\tau _{f}+T_{rev}/3,$ (f) $\tau =\tau _{f}+T_{rev}/4,$ (g) $\tau
=1.8\tau _{f}+T_{rev}/2,$ (h) $\tau =1.8\tau _{f}+T_{rev}/3,$ and (i) $\tau
=1.8\tau _{f}+T_{rev}/4,$ respectively.

\bigskip

Fig. 2. Classical map representing the final angle $\theta $ as a function
of initial angle $\theta _{0}$ for (a) $\tau =0.5\tau _{f},$ (b)$\tau =\tau
_{f},$ (c) $\tau =3\tau _{f},$ and (d) $\tau =10\tau _{f}$.

\bigskip

Fig. 3. Accumulative angular squeezing. Graphs are shown in double
logarithmic scale.

\bigskip

Fig. 4. Spatial distribution of a classical ensemble of atoms after a single 
$\delta $-kick. In Figs. (a) and (b), the initial temperature, $T$ of the
ensemble is zero. In (c) and (d), $k_B T = (1/9) k_l^2
[\int_{-\infty}^{\infty }V(t)dt]^2/m$, where $V(t)$ is the depth of the
potential produced by the standing wave, and $k_B$ is Boltzmann constant.
Figures (a) and (c) show the spatial distribution at focal time $\tau =\tau
_{f}$, while in figure (b) and (c) $\tau =2.5\tau _{f} $.

\bigskip

Fig. 5. Accumulative squeezing of atoms in a pulsed optical lattice at
finite temperature (classical description). The minimal localization factor
as a function of the kick number is shown in double logarithmic scale. The
solid line corresponds to zero initial temperature. The dashed and dotted
lines correspond to $k_B T = (1/9) k_l^2 [\int_{-\infty}^{\infty
}V(t)dt]^2/m $ and $k_B T = (4/9) k_l^2 [\int_{-\infty}^{\infty }V(t)dt]^2/m$%
, respectively.

\bigskip

Fig. 6. Spatial distribution for the optimized sequence of four $\delta $%
-pulses. The upper row shows the spatial distribution averaged over 100
atomic ensembles (each containing 5000 atoms), the lower figures show the
distribution of atoms in one of the ensembles. In (a), the distribution is
plotted at the time of the second pulse, that is delayed by $\tau =3.02\tau
_{f}$ with \ respect to the first pulse, in (b) - at the time of the
(combined) third and fourth pulses delayed by $\tau =1.35\tau _{f}$ after
the second pulse. Figure (c) shows the distribution of atoms at the time of
the maximal squeezing.

\bigskip

Fig. 7. Time evolution of the angular distribution of a (classical) ensemble
of $\delta $-kicked 3D rotors at finite temperature. The probability of
finding a rotor inside the solid angle element $\sin \theta d\theta d\phi $
is plotted on a sphere. Picture (a) shows the distribution at $\tau =0$
(time of the kick). Figures (b), (c), and (d) correspond to $\tau =\tau _{f}$%
, $\tau =3.3\tau _{f}$ and $\tau =5\tau _{f}$, respectively. The initial
temperature corresponds to $k_B T = (1/100) \mu^2 r^2
[\int_{-\infty}^{\infty} {\cal E}(t) dt]^2/I^2$.

\bigskip

Fig. 8. Contour plot for the time-dependent angular distribution function $%
2\pi \sin (\theta )|\Psi (\theta ,\tau )|^{2}$ of a three-dimensional rotor
(molecule) subject to a strong ''polarization-type'' interaction, Eq.(\ref
{polarization}) with a Gaussian pulsed laser field: ${\cal E}^{2}(t)(\alpha
_{\Vert }-\alpha _{\bot })I/4\hbar ^{2}=3\times 10^{3}\exp [-\left( \tau
/0.01\right) ^{2}].$ Here $\tau =t\hbar /I.$ The molecule resides initially
in the isotropic ground angular state ($J=0,m=0).$ Angular focusing and
rainbows emerging from each of the focal regions can be seen. The angular
rainbows are made of moving rings located symmetrically with respect to the
equatorial plane.

\newpage

\begin{table}[tbp]
\begin{tabular}[t]{|c|c|c|}
\hline
No of kicks & $O_{acc}$ & $O_{opt}$ \\ \hline
2 & 0.33 & 0.31 \\ 
3 & 0.26 & 0.20 \\ 
4 & 0.21 & 0.11 \\ 
5 & 0.18 & 0.07 \\ \hline
\end{tabular}
\caption{}
\end{table}

\end{document}